\documentclass[useAMS, usenatbib]{mn2e} 
\usepackage{aas_macros}
\usepackage{graphicx}
\usepackage{subfig}
\usepackage{amsmath}
\usepackage{mathtools}
\usepackage{pdflscape}
\usepackage{longtable}
\long\def\symbolfootnote[#1]#2{\begingroup%
\def\thefootnote{\fnsymbol{footnote}}\footnote[#1]{#2}\endgroup}

\title{Multi-Dimensional Bayesian Membership Analysis of the Sco OB2 Moving Group.}
\author[A.C. Rizzuto et al.]{A.C. Rizzuto$^{1}\thanks{Present address: Department of Physics and Astronomy, Macquarie University, Sydney, 2109, Australia}$, M.J. Ireland$^{1\star}$, J.G. Robertson$^{1}$\\
$^{1}$Sydney Institute for Astronomy (SIfA), School of Physics, University of Sydney NSW, 2006, Australia}

\newcommand{\mupar}{\mu_{\parallel}}
\newcommand{\muperp}{\mu_{\perp}}
\newcommand{\muparg}{\nu_{\parallel_g}}

\newcommand{\vrad}{v_r}
\newcommand{\vradg}{v_{r_g}}
\newcommand{\vradf}{v_{r_f}}

\newcommand{\muparf}{\mu_{\parallel_f}}
\newcommand{\phig}{\boldsymbol{\phi}_g}
\newcommand{\phif}{\boldsymbol{\phi}_f}
\newcommand{\mualf}{\mu_{\alpha}}
\newcommand{\mudel}{\mu_{\delta}}

\begin{document}

\pagerange{\pageref{firstpage}--\pageref{lastpage}} \pubyear{2011}

\maketitle

\begin{abstract}
We present a new high-mass membership of the nearby Sco OB2 association based on HIPPARCOS positions, proper motions and parallaxes and radial velocities taken from the \citet{kharchenko07} catalogue. The Bayesian membership selection method developed makes no distinction between subgroups of Sco OB2 and utilises linear models in calculation of membership probabilities. We select 436 members, 88 of which are new members not included in previous membership selections. We include the classical non members $\alpha$-Cru and $\beta$-Cru as new members as well as the pre-main-sequence stars HIP 79080 and 79081. We also show that the association is well mixed over distances of 8 degrees on the sky, and hence no determination can be made as to the formation process of the entire association.
\end{abstract}


\section{introduction}
\label{intro}
\begin{figure}
\includegraphics[width=0.5\textwidth]{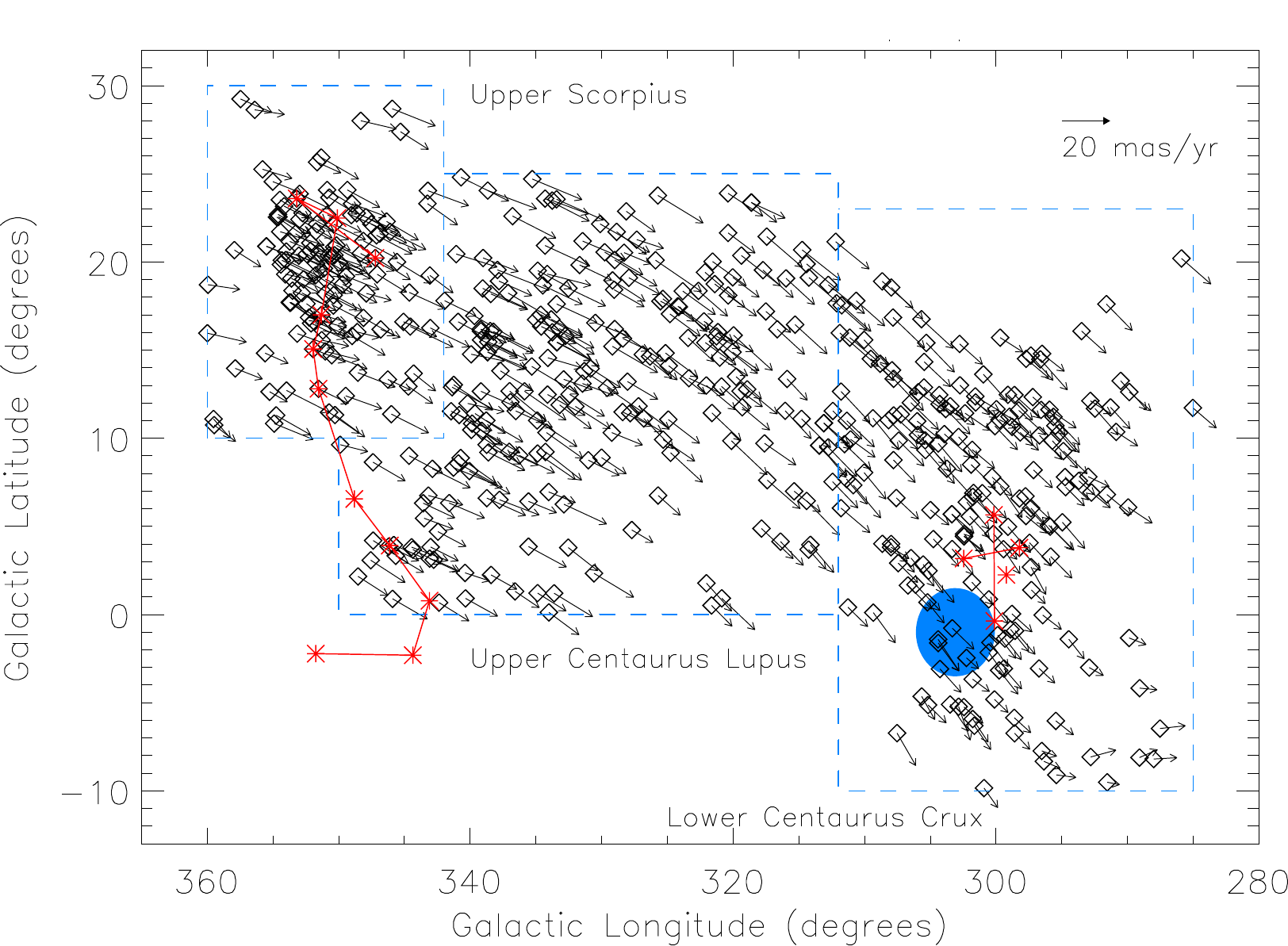}
\caption{Positions of Sco OB2 on the sky with representative HIPPARCOS proper motion vectors. The members shown in this plot are taken from the \citet{zeeuw99} membership 
determination, and show the windows (blue) used for each sub-group membership search. The familiar ``Southern Cross"  and the constellation Scorpius are plotted in red, and the Coalsack nebula is 
shaded in blue.}
\label{zeeuw_sco_cen}
\end{figure}

The Scorpius-Centaurus-Lupus-Crux OB Association (Sco OB2, Sco-Cen) is the nearest location with recent massive star formation. The association was first identified by  \citet{kapteyn14} during an investigation of the parallaxes of 319 bright OB stars in the region of sky occupied by Sco-Cen. Following this, other kinematic studies confirmed that Sco-Cen is indeed a moving group (\citet{plaskett28}, \citet{blaauw46}, \citet{bertiau58}, \citet{petrie62}, \citet{jones71}, \citet{zeeuw99}). Since its discovery, Sco-Cen has been classically divided into three distinct sub-groups (see Figure \ref{zeeuw_sco_cen}), Upper-Scorpius (US), Upper-Centaurus-Lupus (UCL), and Lower-Centaurus-Crux (LCC) \citep{blaauw46}, with mean parallaxes of 6.9, 7.1 and 8.5 milli-arcseconds respectively, or distances of 145, 143 and 118\,pc \citep{zeeuw99}. UCL and LCC have little interstellar material associated with them, whereas filamentary material can be observed towards US which is connected to the Ophiuchus cloud complex, a region of ongoing star formation \citep{geus92}. Photometry has demonstrated that the Ophiuchus cloud complex is on the near side of US at approximately 125\,pc, and isochrone fitting gives ages for the sub-groups as ~5\,Myr for US, ~16\,Myr for UCL and ~16\,Myr for LCC \citep{geus89}. 

US has received significantly more attention than the other two subgroups of Sco-Cen primarily due to its relatively compact size. The age spread of US members is tightly bunched around 5\,Myr, which has led to the conclusion that star formation in US was an externally triggered process \citep{preibisch08}.  The triggering effect has been identified as the shockwave created by several supernova explosions in UCL which occurred approximately 12\,Myr ago. Observations of the kinematics of the large HI loops surrounding Sco-Cen suggest that the shockwaves passed US approximately 5\,Myr ago, which agrees with the stellar age of the subgroup \citep{geus92a}.

UCL and LCC have been subject to less study than US. This is primarily due to their relative lack of concentration on the sky and the closer overlap with the Galactic plane, which makes separation of members and field stars more difficult. Also, large portions of UCL and LCC are not observable from the northern hemisphere (where many early investigations were conducted), and early investigation into pre-main sequence stars focused on dark and reflection nebulae, which are less frequent in UCL and LCC \citep{preibisch08}. Only in the last two decades have true high quality measurements of star motions on the sky (HIPPARCOS) become available, which have allowed separation of UCL and LCC members from the field \citep{zeeuw99}. The formation processes which shaped UCL and LCC are as yet unknown, but are expected to be significantly more complicated than that of US. 

\citet{elmegreen77} propose that OB associations are formed through successive bursts caused by shockwaves and ionisation fronts in molecular clouds. OB stars drive ionisation fronts into neutral material in their vicinity. In dense material the ionisation front is preceded by a shock front as it moves into a nearby molecular cloud. \citet{elmegreen77} suggest that the gas accumulated between the two fronts becomes gravitationally unstable and can collapse to form stars. These new stars will then drive the process further into the remaining molecular cloud. This model agrees with the observations that OB associations tend to display a series of distinct subgroups \citep{ambartsumian55}. It is possible that Sco-Cen formed in this way.

The latest high-mass membership study of Sco-Cen was carried out by \citet{zeeuw99} who identified a total of 521 members. The search combined two different selection methods, the convergent-point and ``Spaghetti'' methods \citep{bruijne99,hoogerwerf99}, which use proper motion and parallax to differentiate between members and non-members in distinct regions of sky corresponding to each Sco-Cen subgroup. The convergent-point method has long been a means by which to identify moving group members \citep{jones71}, though it does have some clear weaknesses. The method is one-dimensional, with the selection of member stars being based only on proper-motion in a specific direction. Additionally, the convergent-point method has a bias towards selecting more distant stars and stars with small proper motion. The "Spaghetti" method is stronger than the convergent-point method in that it takes into account both proper motion and parallax, though it does not constrain the radial velocity of members. This indicates that the development of a selection method which makes use of radial velocity is a clear avenue for improvement.

The Sco OB2 association is a useful and readily available astrophysical laboratory. It is the largest group of newly formed stars in close proximity to the sun and provides an ideal testing ground for new star formation and evolution models. Despite, this Sco-Cen is still relatively poorly studied when compared to other associations such as the Pleiedes, which is only a fraction of the size of Sco-Cen.

\section{Multi-Dimensional Bayesian Membership Selection}
\label{method}

\begin{figure*}
\subfloat[Velocity out of Galactic Centre.]{\label{linearu}\includegraphics[width=0.45\textwidth]{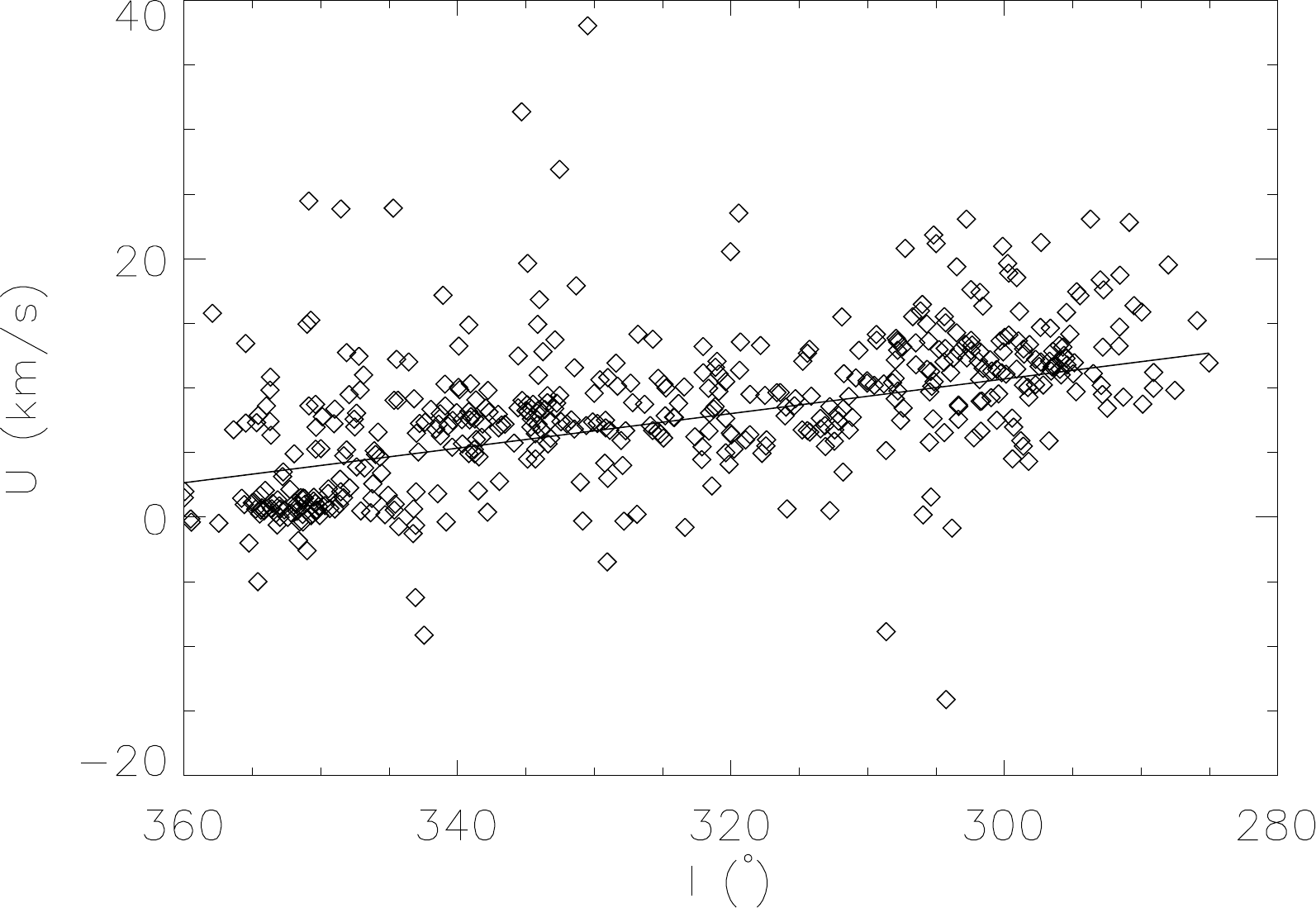}}
\subfloat[Velocity along Galactic Rotation.]{\label{linearv}\includegraphics[width=0.45\textwidth]{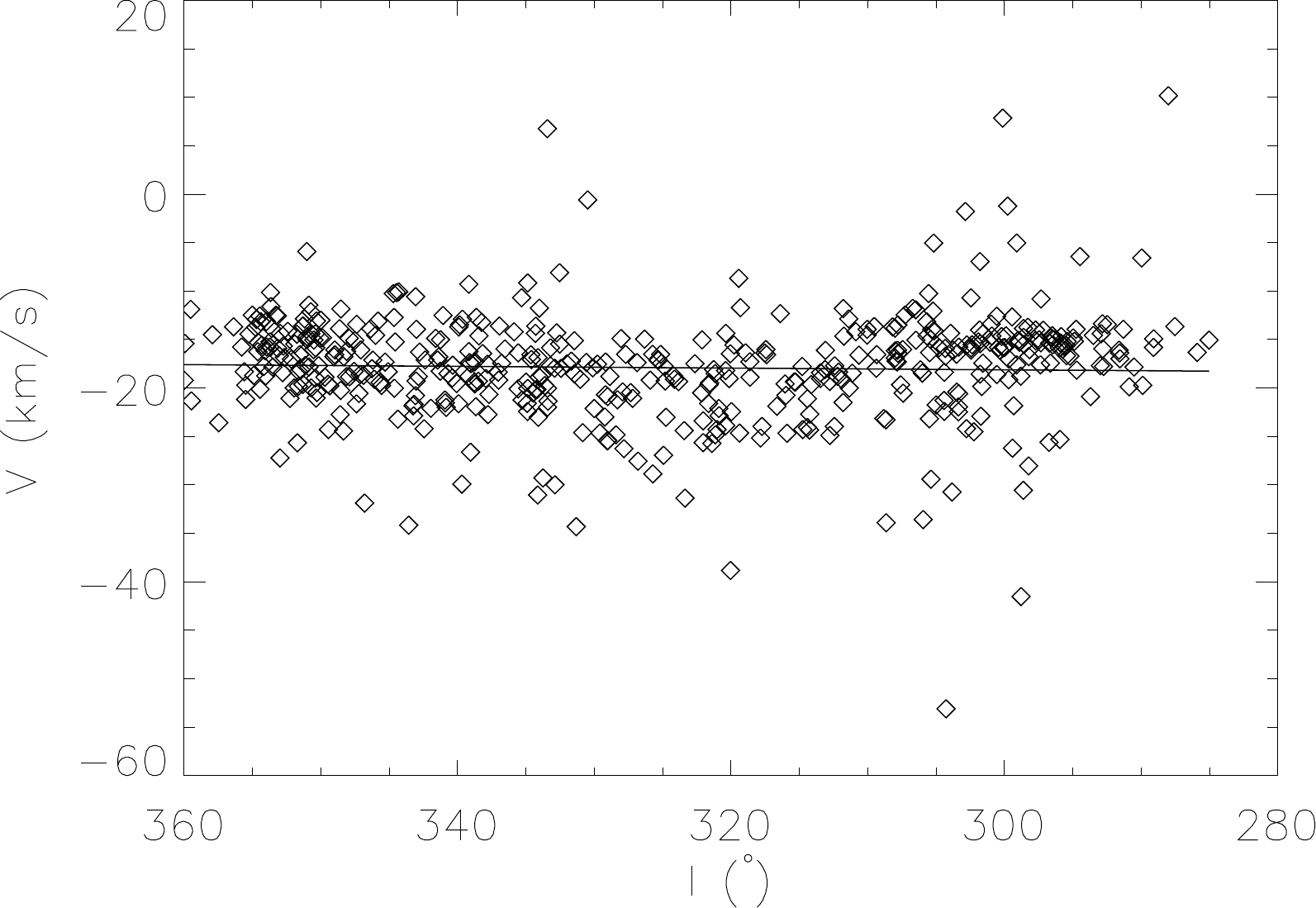}}\\
\subfloat[Velocity out of Galactic Plane.]{\label{linearw}\includegraphics[width=0.45\textwidth]{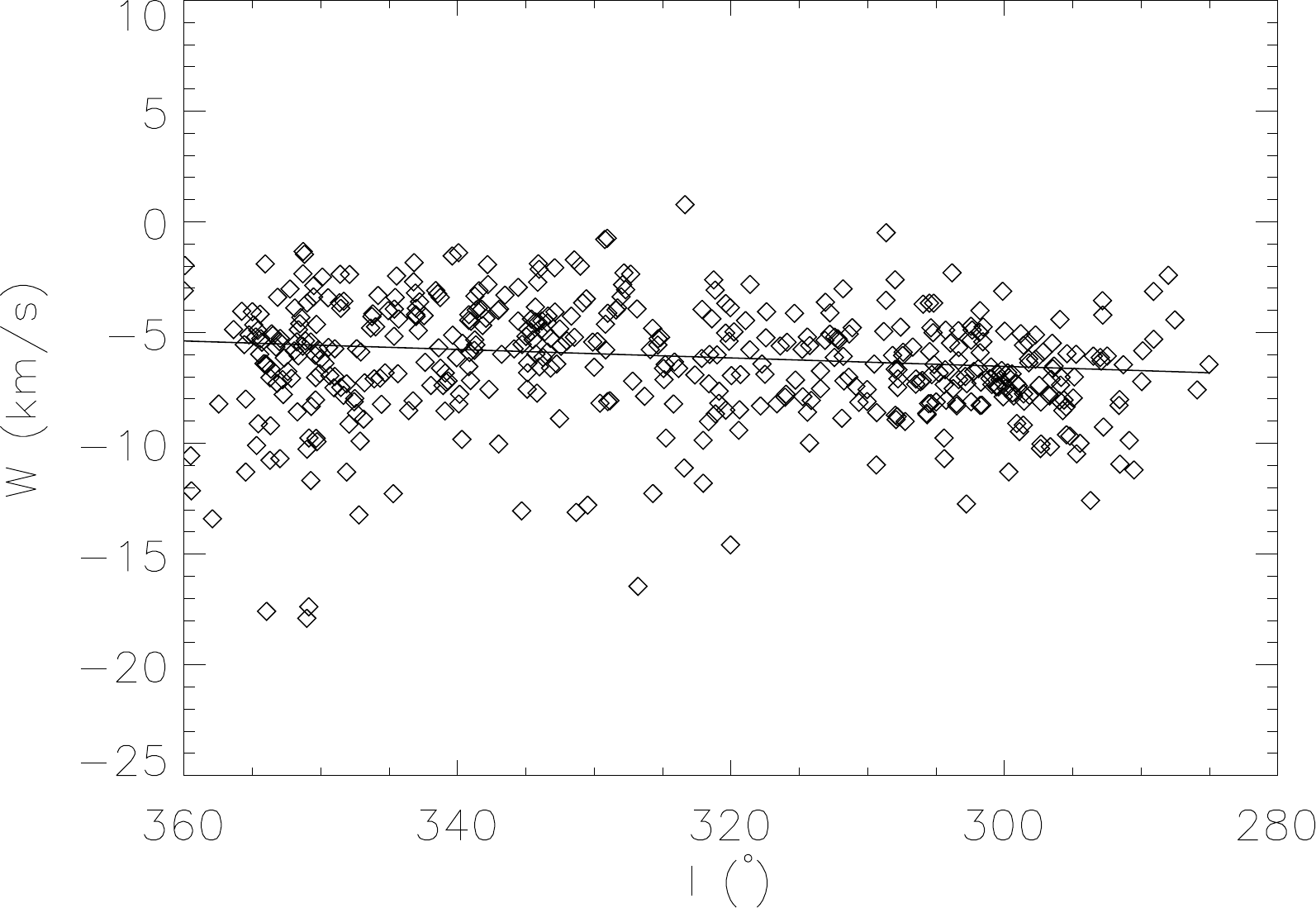}}
\subfloat[Distance]{\label{distline}\includegraphics[width=0.45\textwidth]{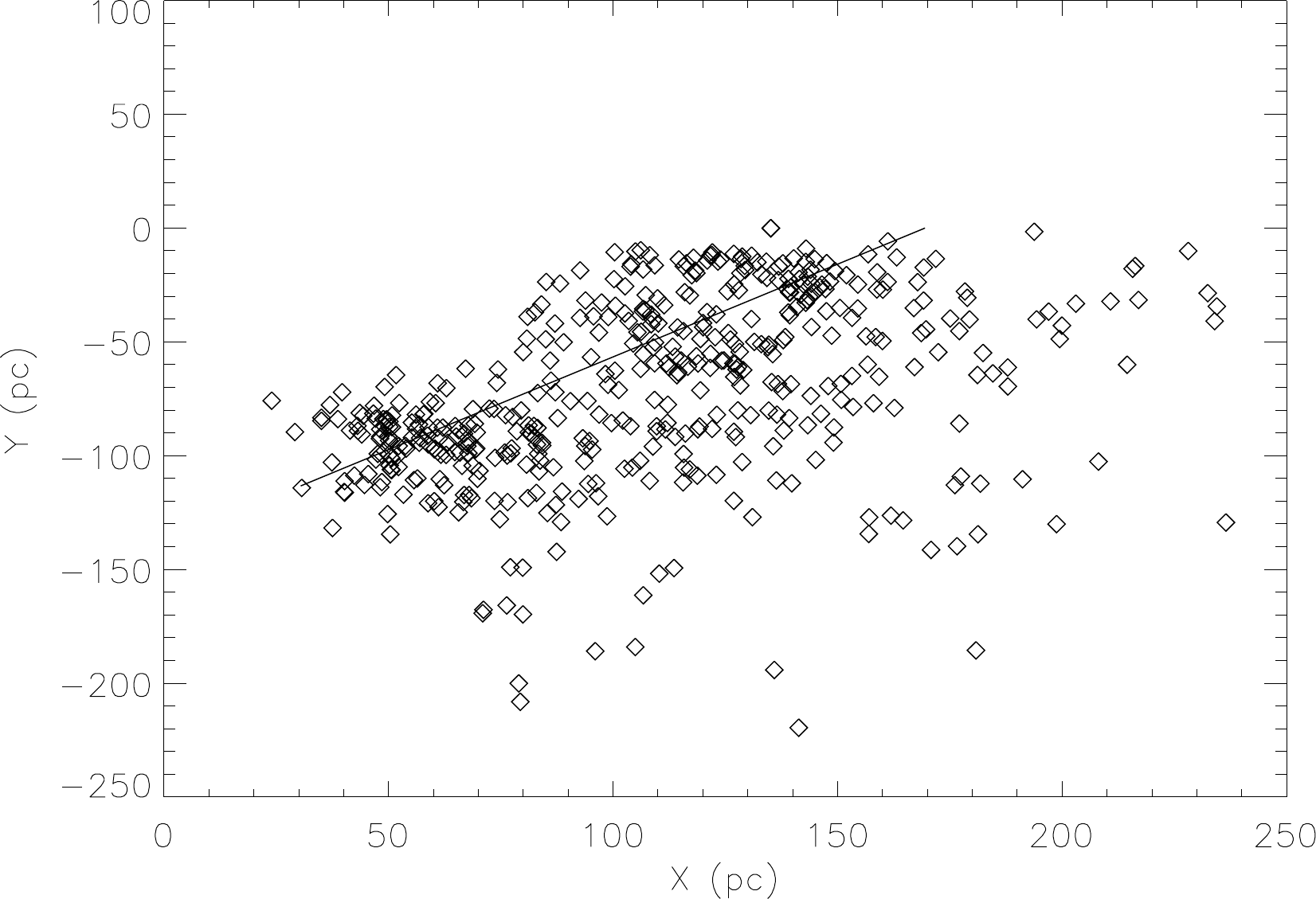}}\\
\centering
\subfloat[Galactic latitude and spread.]{\label{linearb}\includegraphics[width=0.45\textwidth]{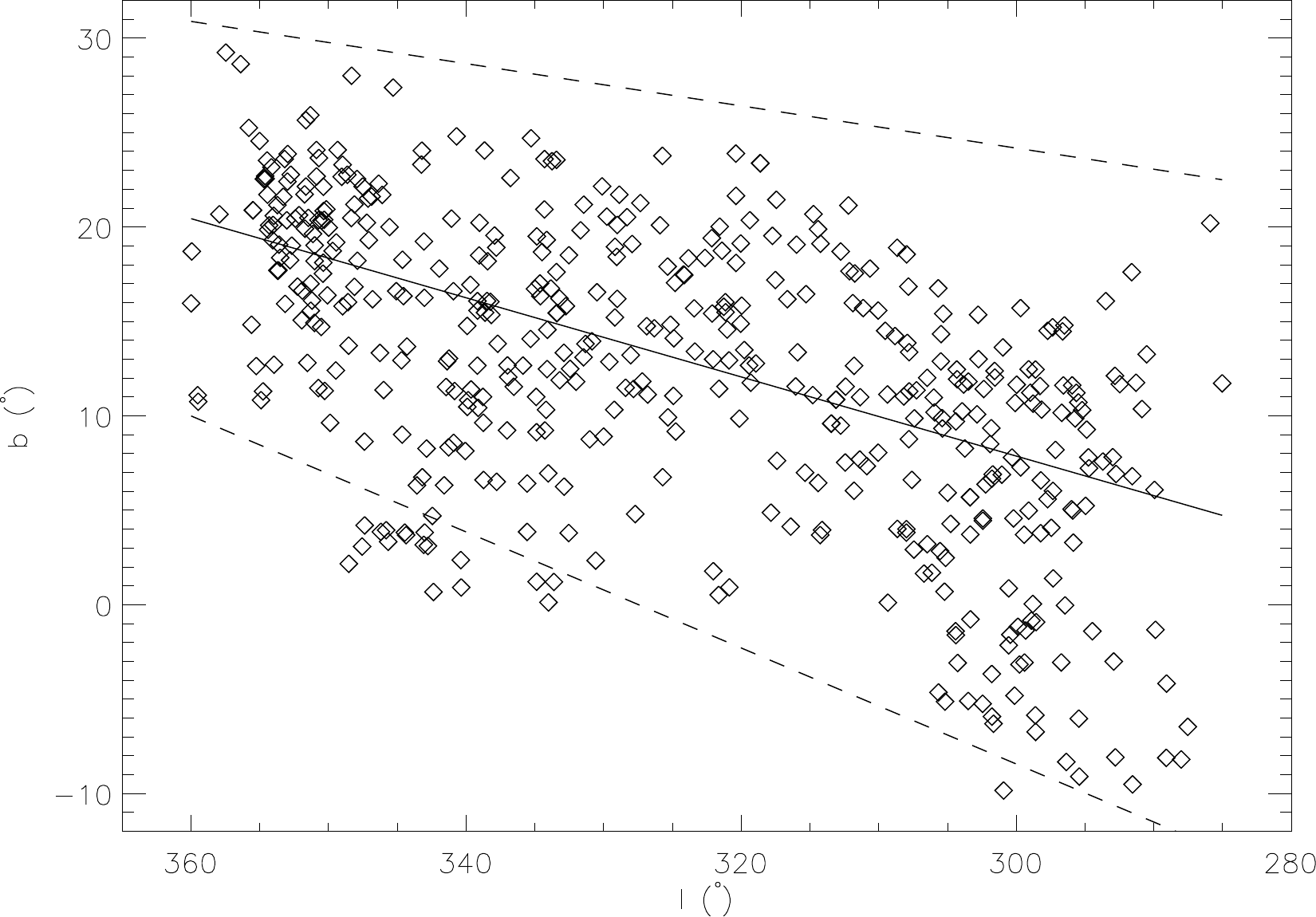}}
\caption{ The Galactic velocity components, distance and Galactic latitude of the \citet{zeeuw99} members. The linear trends with Galactic longitude are shown on each graph. Figure \ref{linearb} shows both the linear trend in Galactic latitude and the linear trend in the spread of Galactic latitude ($\sigma_{b}$) as the dotted lines above and below the solid line. The origin in Figure \ref{distline} is the position of the sun, and stars with a greater distance have larger distance errors.}
\label{linearstuff}
\end{figure*}

Since the last membership selection of \citet{zeeuw99}, there has been a vast improvement in the
available astrometric data for stars in the Sco OB2 field of the sky. The HIPPARCOS \citep{perryman97} mission data has been
reduced in a more accurate way by \citet{leeuwen07}, and catalogues of radial velocities for a large number of the Sco-Cen potential members have become available \citep{kharchenko07}.  These improvements allow one to develop a membership criterion based not only on more accurate proper motion but also parallax, radial velocity and galactic latitude. 

  In the context of the membership analysis described in this paper, the HIPPARCOS data used was that from \citet{leeuwen07} of stars with $B-V \leq 0.6$, within the area of sky bounded by $(285 \leq l \leq 360)$ degrees and $(-10 \leq b \leq 60)$. These are the bright, higher-mass objects in the area of sky occupied by the Sco OB2 association. The radial velocities were taken from the  \citet{kharchenko07} catalogue (CRVAD-2) which is a cross-linking of the All-Sky Compiled Catalogue of 2.5 million stars \citep{kharchenko01}, the General Catalogue of Radial Velocities \citep{barbier00} and a number of other, smaller radial velocity catalogues \citep{kharchenko07}. Measurements from the RAdial Velocity Experiment (RAVE) \citep{ravedr1} are not included in the CRVAD-2 catalogue, and as such, CRVAD-2 can be considered an up-to-date catalogue of stellar radial velocities in the pre-RAVE era. Given that RAVE observations of the Sco OB2 area of sky are not available, CRVAD-2 is the most complete source of radial velocities available for this work. Cross-checking was done between the HIPPARCOS and CRVAD-2 catalogues to find radial velocities for as many of the HIPPARCOS objects as possible. This resulted in 35\% of objects having a known radial velocity, with typical uncertainties of 1 to 5 km~s$^{-1}$. Objects without a radial velocity were assigned a random radial velocity with an extremely large uncertainty, i.e. the radial velocity is treated as unconstrained in the Bayesian membership analysis described below.

Our membership selection uses the six physical parameters of each candidate star to decide upon a membership probability. These parameters are the three Galactic velocity components, and the three position parameters,  distance, Galactic longitude and  Galactic latitude. In our new membership selection, a member must have velocity, Galactic latitude and distance consistent with the five parameters derived from models described below. The five parameters derived from the models $(U, V, W, r$ and $b$) will all be functions of Galactic longitude. 

The Galactic longitude constraint on membership is a simple cut at each end of the association. The Galactic longitude of members is cut at $l=360$ degrees at the US end of the association and at $l=285$ degrees at the LCC end. At the US end, active star formation is in play so it is a logical end-point for the association. At the LCC end, membership probabilities are typically lower because of the confusion with the galactic plane, and the lack of very bright high-mass stars at lower galactic longitudes means there is no clear reason to extend our search further. There may of course still be associated recent star formation beyond LCC, but examining the possibility of this is beyond the scope of this study. In the following subsections we will describe our model for the association, the computational framework and the Bayesian membership selection.

\subsection{Linear Models}
 It is important to first note that the Galactic velocities of the \citet{zeeuw99} members show a clear linear trend with Galactic longitude, with the trend being most pronounced in the component of the velocity pointing out of the Galactic Centre (U) (see Fig. \ref{linearu}). This trend is also present in the distance of group members, Galactic latitude and the spread of Galactic latitude of \citet{zeeuw99} members. Note that the linear trend in the spread of latitudes was derived by assuming that the window sizes defining the subgroups (see boxes in Figure \ref{zeeuw_sco_cen}) were a two-sigma coverage of the Sco-Cen association. This produced a standard deviation of Galactic latitude for each subgroup. A linear trend against Galactic longitude was then fitted to these three values. We denote the Galactic latitude standard deviation trend as $\sigma_b$. 

We characterise the distance of the group members by a line in 3D space,which removes the anthropocentric bias associated with a simpler model where parallax is a function of galactic longitude . The line fitted to the distance of the \citet{zeeuw99} group members is of the form;
\begin{equation}
r = (A\cos{l} + B\sin{l})^{-1},
\label{linear_dist}
\end{equation}
where A and B are the fitted parameters and $l$ is the Galactic longitude. Figure \ref{distline} displays this fitted line and the \citet{zeeuw99} members. Note that the points with larger distances also have proportionally larger errors, and so these outlier points did not contribute to the fit in any significant way.

The trends span all three subgroups (US, UCL and LCC) suggesting that perhaps Sco OB2 can be modelled as a continuum rather than three separate subgroups, with the galactic velocity components, distance and Galactic latitude following a linear trend. The linear trend fitted to the \citet{zeeuw99} results will later be used in the calculation of membership probabilities. 

The $\tt{IDL}$ software package $\tt{mpfit}$ was used to obtain the parameters of the linear trends for the five parameter space variables. The fitted parameters can be seen in Table \ref{paramstable}. Our choice of fitting a linear trend was based on there being no clear observational distinction between the subgroups, and is not a claim of continuous star formation.

\begin{table}
\centering
\begin{tabular}{c c c}
\hline\\
 & Slope & Intercept $(l=0)$ \\
\hline\\
$U$ & -0.13 & 50.9 \\
$V$ & -0.009 & -20.8 \\
$W$ & -0.019 & -12.3 \\
$b$ & 0.21 &-55.0 \\
$\sigma_{b}$ & -0.01 & 45.6 \\
            & A  &  B \\
$r$ &  0.0059 & -0.0072\\
\hline
\end{tabular}
\caption{The linear model parameters derived for the \citet{zeeuw99} members which were used. The galactic velocity ($U$, $V$, $W$) parameters are in units of km~s$^{-1}$, the galactic latitude ($b$, $\sigma_{b}$) parameters are in degrees and the distance (r) parameters are in pc$^{-1}$.}
\label{paramstable}
\end{table}

\subsection{The Convergent-Point and Expected Values}
\subsubsection{The Convergent-Point}
 The membership selection method we have developed makes use of the concept of the convergent-point which was first introduced by \citet{jones71} and modernised by \citet{bruijne99}. The convergent-point is the point on the sky at which a group of stars with a common Galactic velocity will appear to converge to and has formed the basis of one of the search methods used to populate the current membership of Sco-Cen \citep{zeeuw99}. The underlying principle of our new membership selection and its relation to the convergent-point is as follows: For each candidate member, expected Galactic velocity components $(U_g(l), V_g(l), W_g(l))$, distance ($r_g$) and Galactic latitude ($b_g$) can be calculated from the linear trends as functions of the Galactic longitude ($l$) of the candidate. From the expected Galactic velocities, a convergent-point is calculated $(l_{cp},b_{cp})$ and based on this convergent-point, expected proper motions and radial velocity can be found (described below). Hence, for a given star to be a considered a member, it is required to have proper motions, radial velocity and distance which are consistent with the expected values calculated from the linear models in Galactic longitude.

The significance of the convergent-point to the membership selection developed in this paper relates to the velocity coordinate system of the data. The velocities provided by the HIPPARCOS and CRVAD-2 catalogues are two proper motions (equatorial coordinates) and one radial velocity, while the linear models provide Galactic velocity and parallax as a function of Galactic longitude. The concept of the convergent-point allows a direct and easily computed conversion from the model Galactic velocities to the velocity coordinate system of the available data. The convergent-point can be calculated as follows;

\begin{equation}
\begin{array}{c}
\tan(l_{cp}) = \frac{V_{g}(l)}{-U_{g}(l)},\\
\\
\tan(b_{cp}) = \frac{W_{g}(l)}{\sqrt{(U_{g}^2(l) + V_{g}^2(l))}}, \\
\end{array}
\label{conpoint}
\end{equation}
where $l_{cp}$ and $b_{cp}$ are the coordinates of the convergent-point on the sky and $U_g(l)$, $V_g(l)$, and $W_g(l)$ are the three Galactic velocity components derived from the linear trends and are functions of Galactic longitude. Thus the convergent-point is also a function of Galactic longitude.

\begin{figure}
\includegraphics[width=0.45\textwidth]{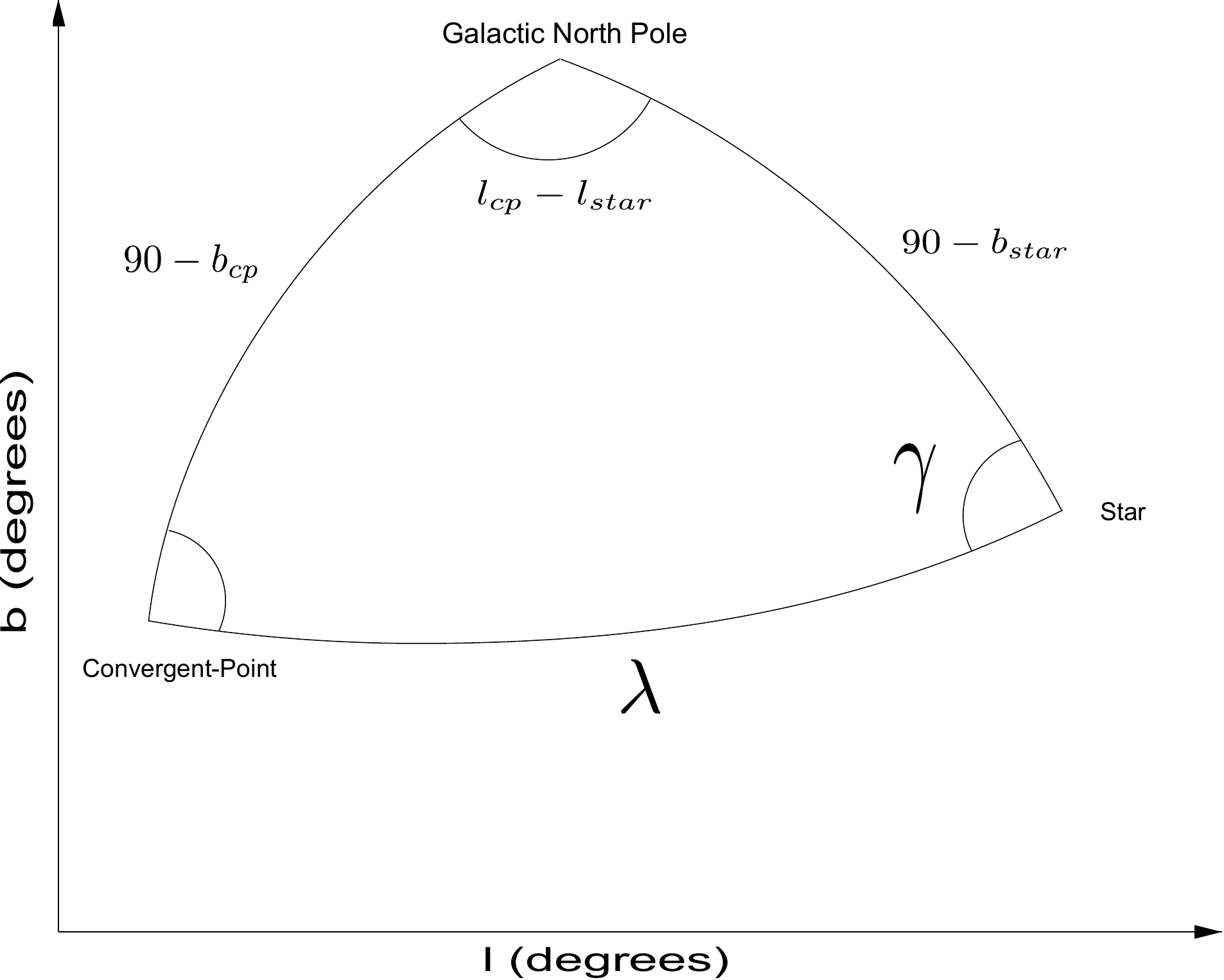}
\caption{The triangle made from the three great-circles joining the star, convergent-point and Galactic north
pole. The angles $\gamma$ and $\lambda$ are the rotation angle and angular distance between the star and the convergent-point respectively. Note that coordinates used here are Galactic coordinates, and also that an analogous diagram can be drawn for equatorial coordinates with the rotation angle $\theta$ and the same angular distance $\lambda$.}
\label{thetriangle}
\end{figure}

Once the convergent-point for a given candidate member is found, the proper motions provided by the HIPPARCOS catalogue are rotated into a new coordinate system with proper motion along the great circle joining the star and the convergent-point and proper motion along the great circle perpendicular to this. The angle of rotation ($\gamma$) is shown in Fig. \ref{thetriangle} (in Galactic coordinates) and is calculated using trigonometric identities. Note that the rotation angle can also be calculated in Equatorial coordinates and will be referred to as $\theta$. The Equatorial rotation angle $\theta$ was used to rotate the HIPPARCOS proper motions into the new coordinate system described above. The rotation is given by the following equation;

\begin{equation}
\begin{pmatrix}
\sin{\theta} & \cos{\theta}\\
-\cos{\theta} & \sin{\theta}
\end{pmatrix}
\begin{pmatrix*}
\mu_{\alpha}\\
\mu_{\delta}
\end{pmatrix*}
=
\begin{pmatrix*}[r]
\mupar\\
\muperp
\end{pmatrix*}
\end{equation}

where $\mualf$ and $\mudel$ are the two HIPPARCOS proper motions, $\muperp$ is the proper motion perpendicular to the direction of the convergent-point and $\mupar$ is the proper motion towards the convergent-point.

 The use of the convergent-point in our method allows the parameters of the models to be easily converted to the framework of the observational data, rather than manipulating the data into a form comparable to the models. The method also allows uncertainties to be calculated via a simple rotational transformation of the HIPPARCOS covariance matrix rather than complicated conversions to Galactic velocity coordinates. It is also important to stress that the convergent-point is a function of Galactic longitude (although indirectly) and hence each star will have a different convergent-point. Hence, we are not using the convergent-point in the traditional sense as the point at which association members converge on the sky, but as a means of constructing a new coordinates system which allows us to convert from Galactic velocities to proper motions and radial velocity. Furthermore, this convergent-point based coordinate system allows a simple membership criterion for $\muperp$: this quantity should be small for a member. This criterion is the basis for the classical convergent-point selection method used by \citet{jones71}.

\subsubsection{Expected Values}
The basis of this new membership selection is that for a candidate star to be a member it must have velocity, distance and position which are consistent with some expected values. The expected values of the five parameters are derived from the linear model described above, and vary for each candidate star with Galactic longitude. The expected Galactic latitude and parallax are taken directly from the linear models, while the expected proper motions and radial velocity require conversion from the linear models for Galactic velocity components $(U_g(l),V_g(l),W_g(l))$. 

To convert the expected Galactic velocity from the linear models, unit vectors in the direction of $\mupar$ and the radial velocity of each star are needed. These will be denoted $\hat{\mathbf{p}}$ and $\hat{\mathbf{r}}$ respectively and can be calculated using spherical trigonometry;

\begin{equation}
\begin{array}{c}
\hat{\mathbf{p}} = \cos{\gamma}\hat{\mathbf{b}} + \sin{\gamma}\hat{\mathbf{l}}, \\
\hat{\mathbf{r}}= (\cos{l}\sin{b},\sin{l}\sin{b},\cos{l}),
\end{array}
\end{equation}

where $\hat{\mathbf{l}}$ and $\hat{\mathbf{b}}$ are the Galactic longitude ($l$) and Galactic latitude ($b$) unit vectors. The expected Galactic velocity $(\tilde{u}_{g}(l))$ is then projected in the direction of these two unit vectors yielding the expected velocity along the great circle joining the convergent-point and the star, and the expected radial velocity, denoted $\muparg$ and $\vradg$. 

\begin{equation}
\begin{array}{c}
\muparg = \hat{\mathbf{p}}\cdot\tilde{u}_{g}(l),\\
\\
\vradg = \hat{\mathbf{r}}\cdot\tilde{u}_{g}(l).\\
\end{array}
\label{projections}
\end{equation}

In order to compare the expected parallel velocity ($\muparg$) to the proper-motion in the direction of the convergent-point for a given star, it is required that $\muparg$ be converted into proper-motion
units. This was done by multiplying by the parallax($\Pi$) and a 
 $M=4.74$ km\,yr\,s$^{-1}$  which is the ratio of the number of seconds in a Julian year and one astronomical unit in kilometres, yielding the expected parallel proper-motion for the given star. This allows ideal membership to be described by Equation \eqref{mcond}. Note that the expected quantities will vary depending on the position of the star in question, and so will be different for each possible member.

\begin{equation}
\left\{
\begin{array}{c}
\muperp=0  \\
\\
\mupar=\Pi M \muparg(l_{cp},b_{cp})  \\
\\
\vrad = \vradg(l_{cp},b_{cp}) \\
\\
r = r_g(l) \\
\\
b = b_g(l)\\
\end{array}\right\}
\label{mcond}
\end{equation}

The above five conditions describe the expected parameters of an ideal member star at a given Galactic longitude. It is expected that there is a velocity dispersion around the expected Galactic velocity of each star due to gravitational interaction of Sco-Cen members. The magnitude of the velocity dispersion used in this selection method was 3\,km~s$^{-1}$ \citep{kraus08}.


\subsection{Preliminary Exclusions}

Using the expected velocities expressed above and the errors associated with the Hipparcos and radial velocity data, a membership probability can be calculated for each star, using Bayes' factors and Monte-Carlo integration. Before the probabilities are calculated, a simple and preliminary yes/no selection was made to initially reject stars which would have low membership probabilities. This was done by finding the squared, inverse-variance weighted difference between each velocity component and its respective expected velocity (Equation \eqref{thethree}), and using a three degree-of-freedom $\chi^2$ probability density function to calculate the probability, $\eta$, of the sum being greater than its calculated value (Equation \eqref{chose}). 
Stars for which a radial velocity was not available in the Karchenko catalogue were treated with a two degree-of-freedom $\chi^2$ probability density function, with the radial velocity term $R = 0$ (see Equation \eqref{thethree}).
Ideally, model likelihood ratios would be calculated for each star in the field, however, stars with a large probability $\eta$ will have a small Bayes' factor and will be excluded by the Bayesian process anyway. 

\begin{equation}
\begin{array}{c}
T = \frac{\muperp^2}{\sigma_{\muperp}^2+\sigma_{int_*}^2},  \\
\\
P = \frac{(\mupar - M\muparg\Pi)^2}{\sigma_{\mupar}^2 + (M\sigma_{\Pi}\muparg)^2 +\sigma_{int_*}^2 },  \\
\\
R = \frac{(\vrad-\vradg)^2}{\sigma_{\vrad}^2 + \sigma_{int}^2},  \\
\end{array}
\label{thethree}
\end{equation}  

\begin{equation}
\eta =  \frac{1}{2^{\frac{3}{2}}\Gamma(\frac{3}{2})}\int_{T+P+R}^{\infty} x^{\frac{1}{2}}e^{-\frac{x}{2}} \mathrm{d}x.
\label{chose}
\end{equation}

The cut off value used for $\eta$ for stars which progressed to the Bayesian analysis  was determined by testing whether setting $\eta$ lower would change the number of stars chosen by the Bayesian analysis at a 50\% threshold. A value of 0.8 was used for $\eta$ , which was well above the condition described, but still able to significantly reduce the time required for computation. Within Sco-Cen, there is a common velocity among members and a dispersion away from the common velocity caused by gravitational interaction.  In Equation \eqref{thethree} we have included internal velocity dispersion $\sigma_{int}$ as a source of deviation from the expected velocities. Conversion to proper motion units is required for $\mupar$ and $\muperp$ using the conversion factor $M=4.74$ km\, yr\,s$^{-1}$ mentioned above and the mean parallax of the \citet{zeeuw99} members. The converted velocity dispersion is denoted $\sigma_{int_*}$.

This preliminary selection did not produce membership probabilities, but simply functioned to filter
out stars which could not be members based on their velocity characteristics. This is why we
do not present a comprehensive analysis of the stars selected, though a brief account is as follows.
Approximately four times the number of stars in the \citet{zeeuw99} membership were included
by this preliminary selection (2051 stars), which is expected since it was designed to let through all but the most clear
non-members (5473 excluded). Note that 15 \citet{zeeuw99} members were rejected, almost all of which were due
to a radial velocity inconsistent with membership, again, this is expected as radial velocity was not used
by \citet{zeeuw99}. A further discussion and additional details on these stars will be given in the
following sections.

\subsection{Bayesian Membership Probabilities}
To find a membership probability for each star in the field of Sco OB2, Bayesian model testing was used. Bayes Factors can be used to calculate the likelihood ratio (R) between two models, in the case of this membership selection the models are that a given star is a member of Sco-Cen, denoted $M_g$, and that a star is a member of the field and not a member of Sco-Cen, denoted $M_f$. The probability of a model given the data is $R/(R+1)$. The model likelihood ratio (R) is given by Equation \eqref{modlike} \citep{sivia}, where D denotes the data, the vertical bar denotes``given" and  P denotes probability. 
\begin{equation}
\\
R = \frac{P(M_g|D)}{P(M_f|D)}\\
\\
\label{modlike}
\end{equation}
Using Bayes' Theorem \citep{sivia} the model likelihood ratio can be expressed as:

\begin{equation}
\\
R = \frac{P(M_g)}{P(M_f)}\frac{P(D|M_g)}{P(D|M_f)} = \frac{P(M_g)}{P(M_f)}K,
\\
\label{rrr}
\end{equation}

where K is the Bayes' factor, and can be written as:

\begin{equation}
\\
K = \frac{P(D|M_g)}{P(D|M_f)} = \frac{\int P(\boldsymbol{\phi}_g|M_g)P(D|\boldsymbol{\phi}_g)\, \mathrm{d}\boldsymbol{\phi}_g}{ \int P(\boldsymbol{\phi}_f|M_f)P(D|\boldsymbol{\phi}_f)\, \mathrm{d}\boldsymbol{\phi}_f},
\\
\label{bayesfact}
\end{equation}

 where $\phig$ and $\phif$ are a set of parameters $\{U, V, W, r, b\}$.



The components which make up the five dimensional integrals in equation \eqref{bayesfact} can be separated into three uncoupled integrals;

\begin{equation}
\begin{array}{l c l}
P(D|M_g)&=&\int_\infty^\infty P(b_g|M_g)P(D|b_g)\mathrm{d}b_g\\
\\
&&\int_\infty^\infty P(r_g|M_g)P(D|r_g)\mathrm{d}r_g \\
\\
&&\int P(\{\mathbf{U}\}|M_g)P(D|\{\mathbf{U}\})\mathrm{d}\{\mathbf{U}\}
\end{array}
\label{separate_them}
\end{equation} 

where $\{\mathbf{U}\}$ is the set of three group Galactic velocity components $\{U_g,V_g,W_g\}$. These cannot be separated because the data do not constrain these components individually. Note that a completely analogous equation can be written for the field integral (denominator of equation \eqref{bayesfact}). Firstly, in the Galactic latitude ($b$) integral, the term $P(D|b_g)$ is the delta function $\delta(b-b_g)$ because the uncertainties in the HIPPARCOS positions are trivially small. Hence the Galactic latitude integral is analytically computed to be simply $P(b|M_g)$, which is a Gaussian with mean and standard deviation taken from the linear models, and truncated at the edge of our search area ($-10< b < 60$). Note that since the mean and standard deviation are function of Galactic latitude, the Gaussian will be different for each star. Similarly, for the field, $ P(D|M_f)$ is a Gaussian with a mean of zero and standard  deviation taken from the spread of Galactic latitudes in the HIPPARCOS catalogue. Again, this was truncated at the edges of our search area.

A more complicated issue is that of the distance $(r)$ integrals. It is important to note that there is a population effect in the HIPPARCOS catalogue towards more distant stars. This is expected as there are intrinsically more stars per unit distance from the sun. This effect will thus be intrinsic to both group and field distance distributions. Hence, the field distribution will be defined by this population effect, and the group distribution will be treated as a Gaussian superimposed on the population effect. The probability distribution of the population effect was determined by binning all the HIPPARCOS stars in our search area into 5\,pc bins and calculating the fraction of stars per bin. The population distribution will be denoted $B(r)$. The terms in the distance integral components for the field are then given by the following;

\begin{equation}
P(r_f|M_f) = B(r),
\label{bias_dist}
\end{equation} 

\begin{equation}
P(D|r_f) = \frac{1}{\sqrt{2\pi\sigma_\Pi^2}}\exp{\big(-\frac{(\frac{1}{r_f}-\Pi)^2}{2\sigma_\Pi^2} \big)}.
\label{guassian_data_f}
\end{equation}

The integral of these two components is computed numerically by first defining the 5\,pc distance bins and calculating the fractional number of stars per bin. This produces a set of $r_f$ values and $P(r_f|M_f)$, their weightings (one for each bin). For each of these $r_f$ values $P(D|r_f)$ is calculated, multiplied by it's corresponding weighting and then totalled to yield the integral. The corresponding group distance distributions are defined in a similar way;

\begin{equation}
P(r_g|M_g) = P(r_g|M_f) C \exp{\big(-(\frac{(r_g-\bar{r}_g(l))^2}{2\sigma_{\bar{r}_g}}) \big)},
\label{group_bias}
\end{equation}

\begin{equation}
P(D|r_g) = \frac{1}{\sqrt{2\pi\sigma_\Pi^2}}\exp{\big(-\frac{(\frac{1}{r_g}-\Pi)^2}{2\sigma_\Pi^2} \big)},
\label{guassian_data_g}
\end{equation}

where $\bar{r}_g(l)$ is the distance taken from the linear trend with Galactic longitude, and C is a normalisation constant which is computed numerically. $\sigma_{\bar{r}_g}$ is the standard deviation of the \citet{zeeuw99} members used to fit the linear distance trend. The assumption that the \citet{zeeuw99} members describe the depth of the association well is  valid because distance was not constrained in the \citet{zeeuw99} membership selection. Note that, as described above $P(r_g|M_g)$ is a Gaussian in distance superimposed on the population distribution, and the factor$ P(D|r_g)$ will be different for each star since $\bar{r}_g(l)$ is a function of Galactic longitude $l$. The group distance integral is computed in the same way as the field integral.

Having separated the Galactic longitude and distance components from the integral in Equation \eqref{bayesfact}, the remaining integral is the contributions from the Galactic velocity components. The factors involved in this integral are described below.

 $P(\phig|M_g)$ (The probability of the parameters given the model that a star is in the group) is modelled as the product of three Gaussian distributions and is shown in Equation \eqref{probgaussiang}. The means of the Gaussians are the expected values of the parameters derived from the linear trends in Galactic longitude $(U_g(l), V_g(l)$ and $W_g(l))$. The spread of these parameters around the expected values is only influenced by the internal velocity dispersion of the association, thus the Gaussians have a standard deviation equal to $\sigma_{int}$.

\begin{equation}
\begin{array}{l c l}
  P(\phig|M_g) & \propto &\exp\big(-\big(\frac{(U-U_g(l))^2}{2\sigma_{int}^2}\\
\\
&&+ \frac{(V-V_g(l))^2}{2\sigma_{int}^2}\\
\\
&& +\frac{(W-W_g(l))^2}{2\sigma_{int}^2}\big)\big)
  \label{probgaussiang}
\end{array}
\end{equation}
\begin{equation}
\begin{array}{l c l}
  P(\phif|M_f)& \propto &\exp\big(-\big(\frac{(U-U_f)^2}{2\sigma_{U_f}^2}\\
\\
&& + \frac{(V-V_f)^2}{2\sigma_{V_f}^2}\\
\\
&&  +\frac{(W-W_f)^2}{2\sigma_{W_f}^2}\big)\big)\\
\label{probgaussianf}
\end{array}
\end{equation}

The distribution for the field, $P(\phif|M_f)$ (Equation \eqref{probgaussianf}), is defined in an analogous way. However, the Galactic velocities are taken from the Galactic Thin Disk with values of $(U_f,V_f,W_f)=(9.0,-6.9,-7.0)$\,km~s$^{-1}$ and standard deviations of $\sigma_{U_f}=19.8$, $\sigma_{V_f}=12.8$, and $\sigma_{W_f}=8.0$\,km~s$^{-1}$  for A-type stars \citep{robin03}.

The probability of the data given the parameters from the group, $P(D|\phig)$, is also defined as the product of three Gaussian distributions, though rather than Galactic velocity, the new coordinates calculated from the convergent-point using the projections in equation \eqref{projections} were used, i.e. $\mupar$, $\muperp$ and $\vrad$. 

The final remaining probability distribution to define is $P(D|\phif)$, the probability of the data given the model that the star is in the field. This was also modelled as three Gaussians with expected values of $\muperp$, $\mupar$ and $\vrad$ calculated in the same way as was done for the group. Note that this involves finding a convergent-point for various points in the parameter space. One may ask why the field stars have a convergent-point, and the answer is that the apparent motion of the field stars is dominated by the motion of the sun, and the rotation of the galactic disk, and so a convergent-point is in fact present for a star at rest with respect to the Galactic Disk. However, in this case the convergent point is really an abstract coordinate system used to find $\mupar$ and $\muperp$. Hence expected values are calculated for each star as described in the previous section. These two probability functions are shown below in Equations \eqref{gaussiang} and \eqref{gaussianf}.

\begin{equation}
\begin{array}{l c l}
P(D|\phig)  & \propto &\exp{\big(-\big(\frac{\muperp^2}{2\sigma^{2}_{\muperp}}}\\
\\
& &+\frac{(\mupar-\Pi M\muparg(l)~)^2}{2(\sigma^2_{\mupar}+(M\muparg(l)\sigma_{\Pi})^2)}\\
\\
& &+\frac{(\vrad-v_{r_g}(l)~)^2}{2\sigma^2_{\vrad}}\big)\big)\\
\\
\label{gaussiang}
\end{array}
\end{equation}

\begin{equation}
\begin{array}{l c l}
P(D|\phif)   &\propto &\exp{\big(-\big(\frac{\muperp^2}{2\sigma^2_{\muperp}}}\\
\\
& & +\frac{(\mupar-A\muparf\Pi)^2}{2(\sigma^2_{\mupar}+(A\muparf\sigma_{\Pi})^2)}\\
\\
& &+\frac{(\vrad-\vradf)^2}{2\sigma_{\vrad}^2}\big)\big)\\
\\
\label{gaussianf}
\end{array}
\end{equation}

 The above was implemented using a Monte-Carlo scheme to compute the integrals in equation \eqref{bayesfact}. $10^5$ samples were used for the group integral and $10^6$ were used for the field to adequately sample the larger parameter space of the field distribution. A brief overview of the process is as follows:  From the model distributions ($M_g$ and $M_f$), a set of randomly sampled parameters are taken to form $\phig$ and $\phif$. $M_g$ is in fact a normal distribution in each parameter (U, V, W, r, b), with means taken from the linear trends in Galactic longitude. Hence, $\phig = \{U, V, W, r, b\}$ is a function of Galactic longitude and will be different for each star. From each parameter sample, a convergent-point is calculated and the expected values of $\muperp$, $\mupar$ and $\vrad$ are found with respect to each convergent point. Once this is done, $P(D|\phig)$ and $P(D|\phif)$ can be found and the integral can be calculated. Note that the integral is over the parameters U, V, and W  and not over Galactic longitude, however, these parameters are functions of Galactic longitude for the group integral.

The prior used in the Bayesian selection method, $\frac{P(M_g)}{P(M_f)}$, was generated in an iterative manner. The first prior was the ratio of the number of member stars in the search area to the number of non-member stars in the \citet{zeeuw99} catalogue, with B-V colour less than 0.6 magnitudes. An A-type star with B-V of 0.6 magnitudes at the mean distance of the Sco-Cen association is expected to appear to be 9$^{th}$ magnitude in the visual. Beyond this magnitude the completeness of the HIPPARCOS data greatly declines and so selection of fainter members is not feasible. Subsequent priors were taken as the ratio of the total of the probabilities of membership and the total number of stars in the field. This was repeated until the prior converged (ten iterations). The prior iteration is described in the following equation, with $p_n$ being the membership probabilities, $P_k$ being the $k^{th}$ prior and $N_f$ being the number of stars in the field.

\begin{equation}
P_k = \frac{P(M_g)}{P(M_f)} =\frac{\sum_{n}^{N_f} p_n}{N_f - \sum_{n}^{N_f} p_n },
\label{iterative_prior}
\end{equation}

Once the prior converged, stars with membership probabilities of 50\% or greater were denoted members of the Sco-Cen association, which has been the historical convention.

\section{The New HIPPARCOS Membership of Sco-Cen OB2}
\label{results}
The membership selection method developed in this paper was used to produce an improved membership list for the Sco-Cen OB association. A full list of member stars and their probabilities can be seen in Appendix \ref{bigtable}. Figure \ref{mysco2} displays the selected members on the sky with proper motion vectors. Members of Sco-Cen have now been identified in a much larger area on the sky, with new members at significant distances from the central clustering of the association. While there is clear distinction between Upper-Scorpius and the rest of the association, UCL and LCC now show consistent density and even clumping of members, including in regions that were previously considered sparser boundaries between subgroups. 

\begin{figure*}
\includegraphics[width=0.75\textwidth]{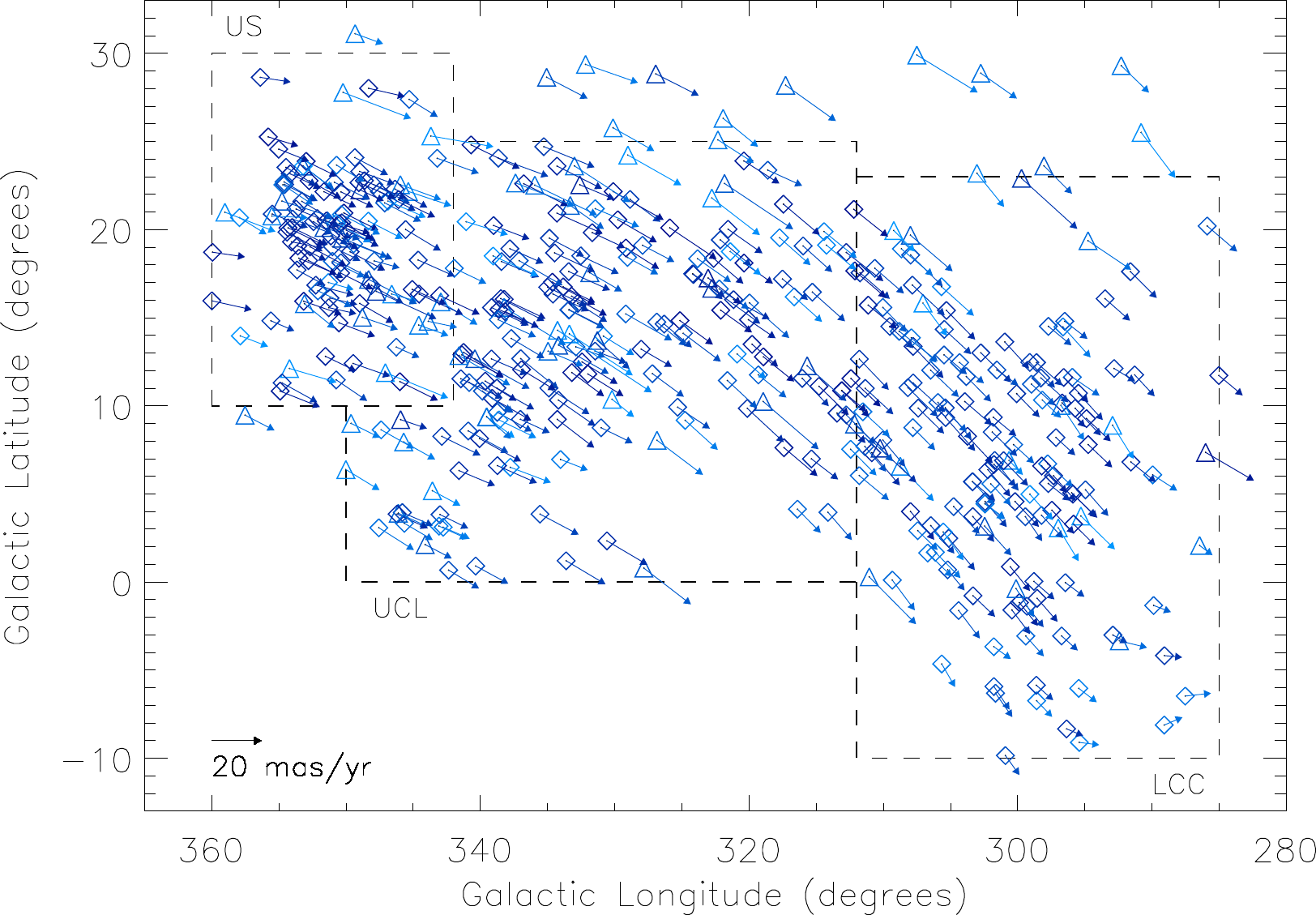}
\caption{Position and proper  motion vectors of the new Sco-Cen members. Note the significantly expanded region of sky within which members have been selected. Diamond shaped points are stars which were included in the last membership \citep{zeeuw99} and triangular points are new members which were not present in the last membership compilation. Darker blue indicates higher membership probability} 
\label{mysco2}
\end{figure*}

\begin{figure*}
\includegraphics[width=0.75\textwidth]{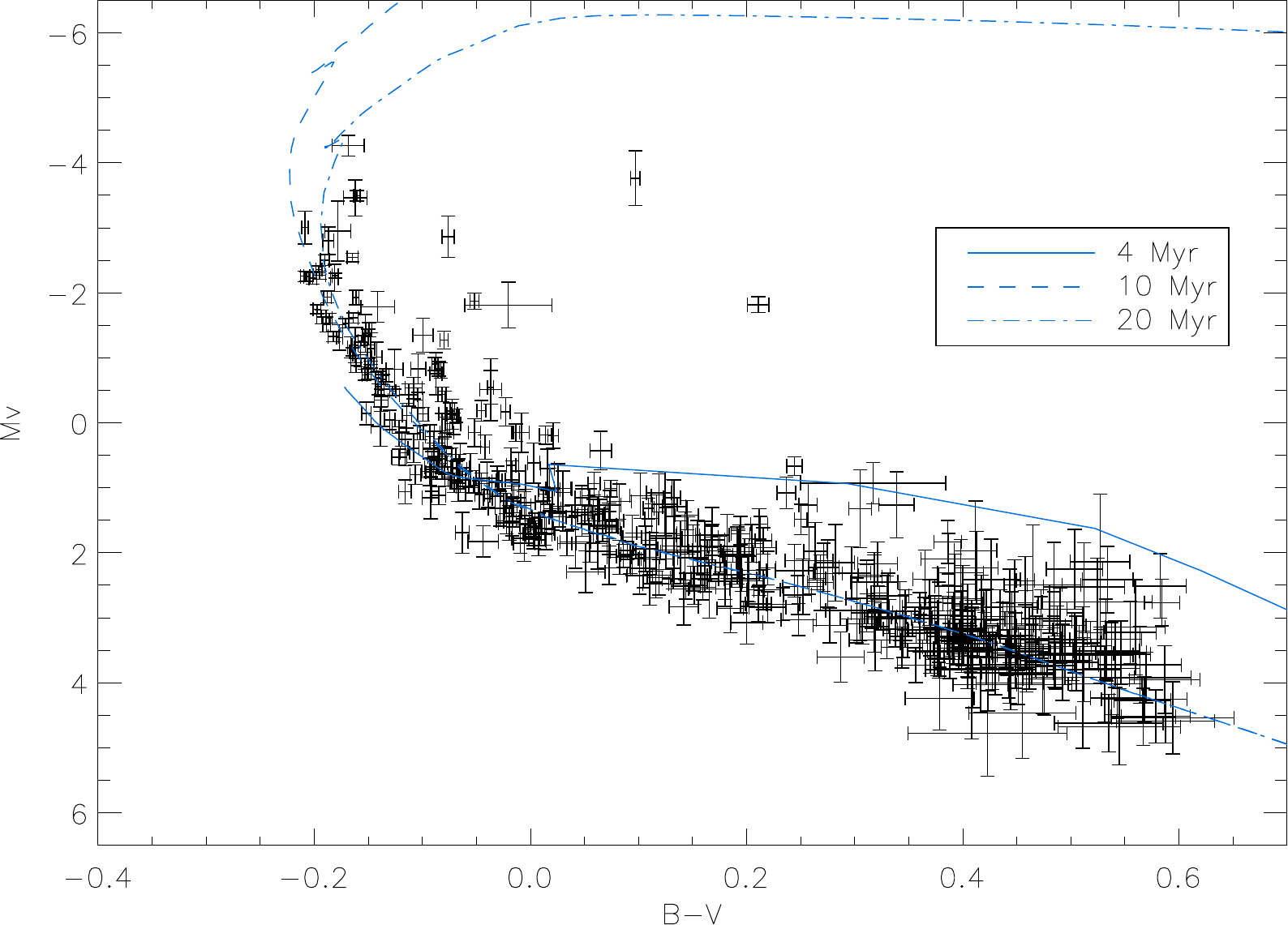}
\caption{Colour-magnitude diagram of the Sco-Cen members selected by the new method with Padova isochrones for 4 (solid), 10 (dashed) and 20 (dotted-dashed)\,Myrs. Note the two isolated stars, these are shifted due to differential reddening caused by the $\rho$-Oph cloud complex which impinges on parts of Upper Scorpius.}
\label{hrdiagram}
\end{figure*}

The selection has produced 436 Sco-Cen members, 348 of which were previous members \citep{zeeuw99} with the rest being new members. 93 \citet{zeeuw99} members have been rejected as members of Sco-Cen, 11 on the basis of parallax, 29 from radial velocity and 53 from HIPPARCOS proper motions. Of the latter 53, only 9 were excluded based on perpendicular proper motion ($\muperp$), which was a major basis of the \citet{zeeuw99} selection. The exclusion of previous members based on HIPPARCOS proper motions, especially the exclusions based on $\muperp$, emphasises the significant effect of the newly reduced data in the membership selection.
Nine of the rejected previous members (HIP 62002, 64661, 68534, 72099, 74449, 77315, 79610, 80208 and 81316) had membership probabilities greater the 40\%, but did not have accurate or available radial velocity measurements. It is likely that these stars can indeed be shown to be members of Sco-Cen with accurate radial velocity measurements.


Importantly, the membership selection has established the classical non-members $\alpha$-Cru and $\beta$-Cru as Sco-Cen stars with membership probabilities of 66 and 73\,\% respectively. This is the first time these stars have been included as members of Sco-Cen. It is important that they are included as they are bright B-type stars with the correct position and parallax and age to have formed as Sco-Cen members. 

Also, we select the classical members $\eta$-Cen and $\mu^{1}$-Sco which were not included in the last membership survey, and we confirm the membership of HIP 59449, as suggested but unconfirmed as a Sco-Cen member \citep{zeeuw99}.

De Zeeuw et al. (1999) suggested, but was unable to confirm that the pre-main-sequence stars HIP 79080 and HIP 79081 were Sco-Cen members. These stars have now been selected as members with probabilities of 51 and 63\,\%. HIP 78094, another pre-main-sequence star, has consistent proper motions for membership but cannot be confirmed as a member due to a poor parallax measurement. We also confirm that
the classical member $\delta$-Sco has proper motions which are inconsistent with group membership. This
star was also excluded from Sco-Cen by de Zeeuw et al. (1999), with the explanation that its binarity, with
period of $\sim$ 20 years could be responsible for producing a proper motion inconsistent with the group.

\section{Discussion}
\label{dis}
The new membership described here does not require the segmentation of Sco-Cen into the three classical sub-groups in order for members to be chosen. Instead Sco-Cen has been treated as a continuum using a linear model to select members and reject non-members. The use of a linear model has removed the need for hard-edged boundaries around selection areas, as seen in the previous membership \citep{zeeuw99}, and has allowed the area in which members can be found to be slightly expanded. Furthermore, the linear trend in the component of the Galactic velocity pointing out of the Galactic Centre (U) demonstrates that the association is indeed expanding. This is expected since OB associations are considered to be loosely bound and is consistent with the idea of a common origin for the entire association.

Figure \ref{hrdiagram} displays photometry taken from the Tycho catalogue \citep{perrymantycho} in a colour-magnitude diagram of the new Sco-Cen members. The Tycho photometry has been converted to standard UBV using the table found in \citet{bessell2000}. We have plotted three Padova isochrones \citep{padova02} at 4, 10 and 20\,Myrs ages which have been reddened to match the Tycho photometry using the following equation \citep{aumer09}.

\begin{equation}
E(B-V) = \left\{
\begin{array}{l l}
0 & \quad d < 70\text{pc} \\
0.47(d-70\text{pc})/1\text{kpc} &\quad d > 70\text{pc}\\
\end{array}\right.
\label{redeqn}
\end{equation}

The value of $E(B-V)$ was calculated to be 0.027 magnitudes and the isochrones were reddened by this amount. 
From the HR diagram, it can be seen that the youngest stars in Sco-Cen are approximately 6\,Myr old, while the older stars appear to be up to 20\,Myr old. This is in agreement with historical estimates of the age of the association \citep{blaauw64a}. Also, Upper Scorpius, the youngest section of Sco-Cen ,is expected to be have formed approximately 5\,Myr ago \citep{preibisch02}, which agrees with the age estimate of the youngest stars in the new membership. Note that the large spread of members around the isochrones in Fig. \ref{hrdiagram} is due to both the parallax measurement error, the effects of unresolved binarity on the photometry and the age spread. There are five stars in Figure \ref{hrdiagram} which appear to be above the main sequence, these stars are listed in Table \ref{redstars}. These five stars, based on their spectral types, are significantly more reddened than the rest of the new association members. This could be due to the impingement of the $\rho$-Oph clouds between the sun and Sco-Cen \citep{geus92}. 

\begin{table}
\caption{List of stars separated from the main sequence in Fig. \ref{hrdiagram}.}
\begin{tabular}{c c c c}
\hline
HIP & $l$ & $b$ &HIPPARCOS Spectral Type\\
\hline
78820&353.2&23.6&B0.5V\\
78933&352.7&22.8&B1V\\
79374&354.6&22.7&B2IV\\
80112&351.3&17.0&B1III\\
80569&357.9&20.7&B2Vne\\
\hline
\\
\end{tabular}
\label{redstars}
\end{table}

\begin{figure}
\includegraphics[width=0.5\textwidth]{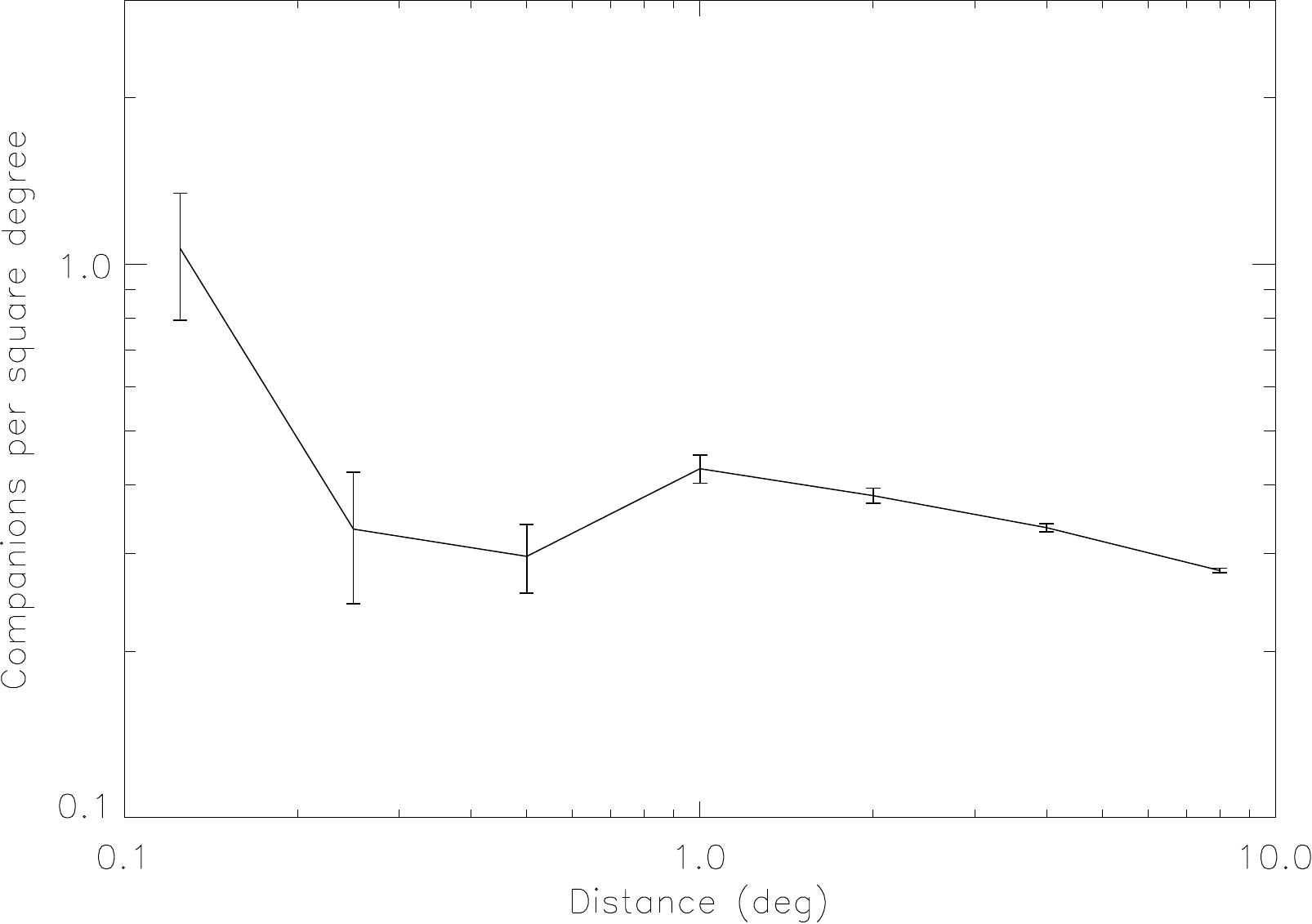}
\caption{A logarithmic histogram of the two point correlation function applied to the new Sco-Cen membership list.}
\label{tpcor}
\end{figure}

A two point correlation function was also applied to the new members of Sco-Cen for the purpose of examining the effects of internal velocity dispersion on mixing the primordial fractal structure of the association. This meant totalling the number of other Sco-Cen stars within certain distance bins. The distance bins were: $(0,0.125)$,$(0.125,0.5)$, $(0.5,1)$, $(1,2)$, $(2,4)$ and $(4,8)$ degrees. Each bin was then area normalised. A histogram of the two point correlation function can be seen in figure \ref{tpcor}. At small separation scales bound binary pairing produces a steep power law \citep{kraus08}. Beyond separations of half a degree, random motions appear to blur out any primordial group structure and produce a shallow slope in the diagram. In this study it is seen that structure on the order of eight degrees is blurred out by the group's velocity dispersion. This implies that hard-edged boundaries between the subgroups of Sco-Cen cannot be defined in a non-arbitrary way. Additionally, this means that it cannot be determined if Sco-Cen formed via the shock front mechanism of \citet{elmegreen77}, as evidence of separated structures has been effectively mixed out.

\subsection{Future Work}
The new membership of Sco-Cen presented here is for the bluer and higher mass population
of Sco-Cen, with a colour $B-V<0.6$. These are the brightest and most well known stars in Sco-Cen. It
has been proposed that there are many thousands of lower mass stars which are members of Sco-Cen but 
which are yet to be identified. The membership selection method developed here can be directly applied to these
lower mass stars once the required data becomes available. The only difficulty is deciding upon the use
of Bayesian prior. In this study, we have used the \citet{zeeuw99} membership as a starting point in defining the prior which resulted in the new membership. There is currently no equivalent for the fainter, lower
mass members. One possibility is to extrapolate the lower mass population based upon the higher mass
population of the new membership in Appendix B, and use this extrapolation as a new prior. This would
involve the use of accurate models to provide a colour-mass relation for young stars. Along a similar
line, spectroscopic techniques can be used to identify lower mass Sco-Cen members. This has been done
for a small region of Sco-Cen by \citet{preibisch02} using the multi-object spectrograph 2dF, but can be greatly
extended with the use of an instrument such as 6dF at the UK Schmidt telescope which is better matched to the large area of sky covered by the association.
As a final point on future membership work, the membership selection method developed here,
with some adaptation, can be applied to other OB associations and moving groups, as was done by \citet{zeeuw99} with their selection method.

Age estimation for the various regions of Sco-Cen has traditionally been achieved through isochrone fitting. However, due to parallax uncertainties and the difficulty in determining the temperature of individual stars accurately, previous age estimates have typically been based on a small number of stars \citep{preibisch02}. Determination of dynamical masses for a number of Sco-Cen targets can allow a more accurate age estimate, and more precise dating across the association. 
\section{Acknowledgements}
This project was supported by the Australian Research Council
(DP0878674). We thank Adam Kraus for helpful discussions on an early
draft, and we thank the anonymous referee for their thoughtful
suggestions.

\appendix

\bibliography{references}
\bibliographystyle{mn2e}
\onecolumn
\section{Membership Table}
\label{bigtable}
\begin{longtable}{c c}
\caption{Membership List and Probabilities} \\
\hline
 HIPPARCOS Number &Probability ($\%$)\\
\hline
\\
\endhead
     50520&            62\\
       50847&            62\\
       52357&            82\\
       52742&            61\\
       53701&            68\\
       53771&            91\\
       54231&            88\\
       54767&            64\\
       55188&            78\\
       55334&            58\\
       55425&            61\\
       55899&            72\\
       56227&            58\\
       56354&            84\\
       56379&            83\\
       56963&            81\\
       57375&            51\\
       57710&            81\\
       57809&            82\\
       57851&            73\\
       57947&            50\\
       57950&            76\\
       58146&            77\\
       58167&            87\\
       58220&            88\\
       58283&            53\\
       58416&            87\\
       58452&            82\\
       58465&            88\\
       58528&            85\\
       58720&            59\\
       58859&            83\\
       58884&            78\\
       58899&            84\\
       58901&            51\\
       59119&            60\\
       59173&            66\\
       59257&            63\\
       59282&            88\\
       59397&            86\\
       59413&            86\\
       59449&            60\\
       59481&            88\\
       59502&            86\\
       59505&            81\\
       59603&            82\\
       59693&            74\\
       59716&            64\\
       59724&            59\\
       59747&            75\\
       59781&            82\\
       59898&            66\\
       59960&            89\\
       60009&            79\\
       60205&            62\\
       60245&            74\\
       60348&            78\\
       60379&            53\\
       60459&            88\\
       60513&            76\\
       60561&            73\\
       60577&            81\\
       60710&            71\\
       60718&            66\\
       60823&            79\\
       60851&            82\\
       60855&            70\\
       61049&            85\\
       61087&            88\\
       61257&            84\\
       61265&            67\\
       61426&            54\\
       61498&            81\\
       61585&            71\\
       61639&            84\\
       61684&            73\\
       61782&            84\\
       61796&            61\\
       62026&            77\\
       62032&            73\\
       62058&            80\\
       62134&            83\\
       62171&            78\\
       62179&            64\\
       62327&            84\\
       62427&            65\\
       62428&            81\\
       62431&            59\\
       62434&            73\\
       62657&            83\\
       62683&            62\\
       62786&            55\\
       63003&            68\\
       63005&            81\\
       63007&            83\\
       63041&            83\\
       63204&            87\\
       63210&            75\\
       63272&            84\\
       63439&            70\\
       63836&            85\\
       63839&            73\\
       63886&            75\\
       63945&            79\\
       63975&            57\\
       64004&            73\\
       64044&            78\\
       64053&            82\\
       64184&            81\\
       64264&            58\\
       64320&            70\\
       64322&            74\\
       64425&            54\\
       64515&            71\\
       64617&            51\\
       64837&            69\\
       64877&            61\\
       64892&            74\\
       64925&            80\\
       64995&            79\\
       65021&            59\\
       65089&            84\\
       65112&            81\\
       65136&            63\\
       65178&            61\\
       65219&            78\\
       65271&            62\\
       65348&            50\\
       65394&            67\\
       65426&            76\\
       65517&            69\\
       65617&            64\\
       65822&            81\\
       65875&            88\\
       65965&            70\\
       66068&            75\\
       66075&            56\\
       66447&            93\\
       66454&            89\\
       66566&            82\\
       66651&            75\\
       66722&            92\\
       66821&            83\\
       66908&            88\\
       67036&            81\\
       67068&            80\\
       67199&            86\\
       67230&            65\\
       67260&            61\\
       67428&            73\\
       67464&            57\\
       67472&            54\\
       67497&            92\\
       67669&            66\\
       67703&            70\\
       67919&            66\\
       67957&            89\\
       67970&            90\\
       67973&            75\\
       68080&            60\\
       68245&            76\\
       68282&            77\\
       68335&            87\\
       68413&            66\\
       68532&            89\\
       68722&            72\\
       68781&            89\\
       68862&            58\\
       68867&            56\\
       69011&            77\\
       69113&            80\\
       69291&            88\\
       69302&            82\\
       69475&            59\\
       69618&            71\\
       69720&            83\\
       69791&            60\\
       69995&            70\\
       70149&            90\\
       70300&            86\\
       70350&            70\\
       70441&            93\\
       70455&            51\\
       70483&            52\\
       70558&            88\\
       70626&            79\\
       70689&            90\\
       70697&            89\\
       70753&            80\\
       70809&            68\\
       70833&            84\\
       70904&            66\\
       70931&            72\\
       70998&            89\\
       71140&            67\\
       71271&            55\\
       71321&            92\\
       71352&            82\\
       71353&            90\\
       71453&            66\\
       71536&            84\\
       71708&            84\\
       71724&            83\\
       71860&            70\\
       71865&            93\\
       72149&            51\\
       72158&            59\\
       72216&            57\\
       72584&            78\\
       72627&            93\\
       72683&            74\\
       72800&            71\\
       72940&            91\\
       73145&            91\\
       73147&            78\\
       73150&            70\\
       73266&            92\\
       73334&            73\\
       73341&            65\\
       73535&            62\\
       73559&            61\\
       73624&            86\\
       73742&            58\\
       73807&            76\\
       73913&            88\\
       73937&            86\\
       73990&            81\\
       74066&            67\\
       74098&            67\\
       74100&            87\\
       74104&            74\\
       74468&            70\\
       74479&            78\\
       74499&            90\\
       74651&            62\\
       74865&            90\\
       74911&            62\\
       74950&            57\\
       74959&            84\\
       74985&            81\\
       75035&            61\\
       75056&            84\\
       75077&            78\\
       75151&            81\\
       75164&            78\\
       75210&            91\\
       75304&            90\\
       75367&            80\\
       75476&            82\\
       75480&            89\\
       75491&            70\\
       75509&            92\\
       75613&            52\\
       75647&            83\\
       75683&            83\\
       75802&            50\\
       75824&            69\\
       75891&            90\\
       75915&            92\\
       76001&            73\\
       76048&            83\\
       76063&            55\\
       76071&            70\\
       76084&            87\\
       76126&            58\\
       76143&            51\\
       76234&            73\\
       76297&            91\\
       76310&            69\\
       76371&            74\\
       76395&            78\\
       76501&            50\\
       76591&            64\\
       76600&            53\\
       76633&            89\\
       76875&            89\\
       76945&            85\\
       76997&            62\\
       77038&            88\\
       77086&            83\\
       77124&            56\\
       77144&            79\\
       77150&            90\\
       77286&            84\\
       77295&            74\\
       77317&            67\\
       77347&            56\\
       77388&            78\\
       77399&            59\\
       77432&            87\\
       77502&            62\\
       77520&            90\\
       77545&            84\\
       77562&            70\\
       77635&            89\\
       77815&            79\\
       77840&            51\\
       77858&            83\\
       77859&            75\\
       77900&            82\\
       77909&            71\\
       77911&            83\\
       77960&            86\\
       77968&            59\\
       78043&            83\\
       78099&            88\\
       78104&            82\\
       78150&            84\\
       78183&            88\\
       78196&            90\\
       78207&            86\\
       78233&            64\\
       78263&            59\\
       78324&            78\\
       78384&            84\\
       78416&            74\\
       78494&            54\\
       78530&            76\\
       78533&            88\\
       78549&            89\\
       78555&            89\\
       78581&            86\\
       78641&            91\\
       78655&            90\\
       78663&            88\\
       78702&            90\\
       78711&            63\\
       78754&            73\\
       78756&            75\\
       78809&            91\\
       78820&            50\\
       78821&            67\\
       78830&            63\\
       78847&            89\\
       78853&            55\\
       78877&            61\\
       78918&            72\\
       78933&            91\\
       78956&            93\\
       78963&            72\\
       78968&            89\\
       78996&            86\\
       79031&            88\\
       79044&            81\\
       79054&            91\\
       79078&            58\\
       79080&            51\\
       79081&            63\\
       79097&            66\\
       79098&            55\\
       79124&            82\\
       79142&            54\\
       79156&            88\\
       79229&            76\\
       79250&            92\\
       79288&            87\\
       79363&            71\\
       79366&            87\\
       79369&            82\\
       79374&            72\\
       79383&            74\\
       79392&            86\\
       79399&            82\\
       79400&            54\\
       79404&            82\\
       79410&            79\\
       79439&            63\\
       79476&            80\\
       79516&            77\\
       79530&            79\\
       79599&            85\\
       79622&            76\\
       79631&            86\\
       79673&            83\\
       79690&            70\\
       79710&            79\\
       79739&            65\\
       79742&            76\\
       79771&            89\\
       79785&            77\\
       79860&            72\\
       79878&            89\\
       79897&            90\\
       79910&            88\\
       79977&            90\\
       80024&            82\\
       80036&            68\\
       80059&            84\\
       80062&            57\\
       80088&            87\\
       80112&            56\\
       80126&            86\\
       80130&            89\\
       80238&            92\\
       80311&            86\\
       80324&            92\\
       80371&            85\\
       80390&            79\\
       80425&            66\\
       80461&            81\\
       80475&            52\\
       80493&            86\\
       80535&            90\\
       80569&            55\\
       80591&            84\\
       80711&            53\\
       80799&            91\\
       80815&            62\\
       80896&            86\\
       80911&            85\\
       81208&            58\\
       81266&            89\\
       81447&            59\\
       81455&            76\\
       81474&            91\\
       81477&            53\\
       81624&            79\\
       81751&            78\\
       81891&            58\\
       81914&            58\\
       81949&            74\\
       81972&            59\\
       82069&            51\\
       82218&            89\\
       82319&            51\\
       82397&            81\\
       82430&            78\\
       82514&            65\\
       82534&            89\\
       82545&            86\\
       82554&            71\\
       82560&            75\\
       82569&            63\\
       82714&            54\\
       83159&            62\\
       83508&            64\\

    \hline
  \label{membershiptab}
\end{longtable}
\twocolumn

\end{document}